\documentstyle[12pt]{article}
\newcommand {\m}{\mu}
\newcommand {\n}{\nu}
\newcommand {\pl}{\partial}

\newcommand {\al}{\alpha}
\newcommand {\be}{\beta}
\newcommand {\ga}{\gamma}
\newcommand {\Ga}{\Gamma}

\newcommand {\la}{\lambda}

\newcommand {\si}{\sigma}

\newcommand {\ep}{\epsilon}

\newcommand {\na}{\nabla}
\newcommand {\del}  {\delta}
\newcommand {\Del}  {\Delta}
\newcommand {\mn}{{\mu\nu}}
\newcommand {\ls}   {{\lambda\sigma}}
\newcommand {\ab}   {{\alpha\beta}}
\newcommand {\half}{ {\frac{1}{2}} }
\newcommand {\fourth} {\frac{1}{4} }
\newcommand {\sqg} {\sqrt{g}}

\newcommand {\Lcal}{{\cal L}}

\newcommand {\intfx} {{\int d^4x}}

\newcommand {\change} {\leftrightarrow}
\newcommand {\ra} {\rightarrow}
\newcommand {\pr}   {{\quad .}}
\newcommand {\com}  {{\quad ,}}
\newcommand {\q}    {\quad}
\newcommand {\qq}   {\quad\quad}

\newcommand {\lb}    {\linebreak}
\newcommand {\nl}    {\newline}
\newcommand {\nn}    {\nonumber}
\newcommand {\ul}    {\underline}
\newcommand {\vs}[1]  { \vspace*{#1 cm} }

\newcounter{eq}
\newcounter{sc}

\def\overleftrightarrow#1{\vbox{\ialign{##\crcr
 $\leftrightarrow$\crcr\noalign{\kern-1pt\nointerlineskip}
 $\hfil\displaystyle{#1}\hfil$\crcr}}}

\begin{document}
\title{    Weak Field Expansion of Gravity and
           Graphical Representation
           \thanks{US-96-03}
                                 }
\author{  
          Shoichi ICHINOSE$^a$
          \thanks{ E-mail address:\ ichinose@momo1.u-shizuoka-ken.ac.jp}
          and Noriaki IKEDA$^b$
          \thanks{ E-mail address:\ nori@kurims.kyoto-u.ac.jp }\\
     $^a$ Department of Physics, University of Shizuoka,       \\
          Yada 52-1, Shizuoka 422, Japan                       \\
     $^b$ Research Institute for Mathematical Sciences, \\
          Kyoto University, Kyoto 606-01,  Japan                  
        }
\date{  August, 1996 }
\maketitle
\setlength{\baselineskip}{0.54cm}
\begin{abstract}
We introduce a graphical representation for a global SO(n) tensor
$\pl_\m\pl_\n h_\ab$, which generally appears 
in the perturbative approach of gravity around the flat space: 
$g_\mn=\del_\mn+h_\mn$. We systematically construct global SO(n) invariants.
Independence and completeness of those invariants
are shown by taking examples of $\pl\pl h$-, 
and $ 
(\pl\pl h)^2$- invariants. They are classified
graphically. Indices which characterize all independent invariants 
(or graphs) are given.
We apply the results to general invariants
with dimension $(Mass)^4$~ and the Gauss-Bonnet identity in 4-dim gravity.
\end{abstract}
\section{Introduction}
\q In n-dimensional Euclidean (Minkowskian) flat space(-time), 
fields are classified 
as scalar, spinor, vector, tensor, ... ,
by the
transformation property under the global SO(n) ( SO(n$-$1,1) ) transformation
of space(-time) coordinates.
\begin{eqnarray}
{x^\m}'=M^\m_{~\n}x^\n\com\q
\label{intro.1}
\end{eqnarray}
where $M$ is a $n\times n$ matrix of SO(n)(SO(n-1,1))
\footnote{
Hereafter we take the Euclidean case for simplicity.
}
.
As for the lower spin fields, the field theory is well defined classically
and quantumly. 

The general curved space is described by the general relativity which is based
on invariance under the general coordinate transformation.
Its infinitesimal form is written as
\begin{eqnarray}
{x^\m}'=x^\m-\ep^\m(x)\com\q |\ep|\ll 1\com\nn\\
\del g_\mn=g_{\m\la}\na_\n\ep^\la+g_{\n\la}\na_\m\ep^\la+O(\ep^2)
=\ep^\la \pl_\la g_\mn +g_{\m\la}\pl_\n\ep^\la+g_{\n\la}\pl_\m\ep^\la
+O(\ep^2)\com\q \label{intro.2}
\end{eqnarray}
where $\ep^\m$ is an infinitesimally-small local free
parameter. 
The general 
invariant composed of purely geometrical quantities and with the mass
dimension $(Mass)^2$ is uniquely given by Riemann scalar curvature, $R$,
defined by
\begin{eqnarray}
\Ga^\la_\mn=\half g^\ls (\pl_\m g_{\si\n}+\pl_\n g_{\si\m}-\pl_\si g_\mn)\com
R^\la_{~\m\n\si}=\pl_\n\Ga^\la_{\m\si}+\Ga^\la_{\tau\n}\Ga^\tau_{\m\si}-
(\n\change\si)\com\nn\\
R_\mn =R^\la_{~\mn\la}\com\q R=g^\mn R_\mn\com\q g=+\mbox{det}g_\mn\pr
                                                      \label{intro.3}
\end{eqnarray}
It is well-known that the general relativity can be constructed purely
within the flat space first by introducing a symmetric second rank tensor
(Fierz-Pauli field) and then by requiring consistency in the field equation
in a perturbative way of the weak field \cite{EA}. 
In the present case, we can obtain the perturbed lagrangian simply by
the perturbation around the flat space.
\begin{eqnarray}
g_\mn=\del_\mn+h_\mn\com\q |h_\mn|\ll 1\pr
                                                      \label{intro.4}
\end{eqnarray}
Then the transformation (\ref{intro.2}) is expressed as
\begin{eqnarray}
\del h_\mn=\pl_\m\ep^\n+h_{\m\la}\pl_\n\ep^\la+\half \ep^\la \pl_\la h_\mn
+\m\change \n +O(\ep^2)\pr
                                                      \label{intro.5}
\end{eqnarray}
In the right-hand side (RHS), there appear 
$h^0$-order terms and $h^1$-order terms.
Therefore the general coordinate transformation (\ref{intro.5}) does not
preserve the weak-field ($h_\mn$) perturbation order.
Riemann scalar curvature is also expanded as
\begin{eqnarray}
& R=\pl^2 h-\pl_\m\pl_\n h_\mn
 -h_\mn (\pl^2 h_\mn-2\pl_\la\pl_\m h_{\n\la}+\pl_\m\pl_\n h) & \nn\\
& +\half \pl_\m h_{\n\la}\cdot \pl_\n h_{\m\la}
-\frac{3}{4}\pl_\m h_{\n\la}\cdot \pl_\m h_{\n\la}
+\pl_\m h_{\m\la}\cdot \pl_\n h_{\n\la}                      & \nn\\
& -\pl_\m h_\mn\cdot\pl_\n h+\fourth\pl_\m h\cdot\pl_\m h
+O(h^3)\com\q                                          & \label{intro.6}\\
& h\equiv h_{\m\m}\pr                                   & \nn
\end{eqnarray}
RHS is expanded into the infinite power series
of $h_\mn$ due to the presence of the 'inverse' field of $g_\mn$,\ 
$g^\mn$, in (\ref{intro.3}).

It is explicitly checked that $R$, defined perturbatively by
the RHS of (\ref{intro.6}), transforms, under (\ref{intro.5}),
as a scalar $\del R(x)=\ep^\la(x)\pl_\la R(x)$, at the order of $O(h)$.
Because the general coordinate symmetry does not preserve the
the weak-field ($h_\mn$) perturbation order, we need
$O(h^2)$ terms in (\ref{intro.6}) in order to verify
$\del R(x)=\ep^\la(x)\pl_\la R(x)$, at the order of $O(h)$.
The first two terms of RHS of (\ref{intro.6}), 
$\pl^2 h$ and $\pl_\m\pl_\n h_\mn$, are two independent global
SO(n) invariants at the order O(h).
We may regard the weak field perturbation using (\ref{intro.4}) as a
sort of 'linear' representation of the general coordinate symmetry,
where all general invariant
quantities are generally expressed by the infinite series of power of $h_\mn$,
and there appears no 'inverse' fields.
One advantage of the linear representation is that the independence
of invariants, as a local function of $x^\m$, 
can be clearly shown because all quantities are written only by 
$h_\mn$ and its derivatives. 
We analyze some basic points of the weak-field expansion and develop
a useful graphical technique.

Mathematically we classify all independent SO(n)-invariants of certain types,
by use of the graph topology.

\newpage
\section{Representation of $\pl\pl h$-tensors and invariants}
\q We represent the 4-th rank global SO(n) tensor (4-tensor),
$\pl_\m\pl_\n h_\ab$, as follows.

\begin{center}
{Fig.1\ 4-tensor $\pl_\m\pl_\n h_\ab$}
\end{center}

\begin{description}
\item[Def 1]\ 
We call dotted lines  {\it suffix-lines}, a rigid line a {\it bond},
a vertex with a crossing mark a {\it h-vertex} and that without it 
a {\it dd-vertex}.
\end{description}
This graph respects all suffix-permutation
symmetries of $\pl_\m\pl_\n h_\ab$\ :\ 
\begin{eqnarray}
\pl_\m\pl_\n h_\ab=\pl_\n\pl_\m h_\ab=\pl_\m\pl_\n h_{\be\al}
\pr\q \label{ddh.1}
\end{eqnarray}
\begin{description}
\item[Def 2]\ 
The suffix {\it contraction} is expressed by connecting the two corresponding 
suffix-lines. 
\end{description}
For example, 2-tensors\ :\ 
$\pl^2h_\ab\ ,\ \pl_\m\pl_\n h_{\al\al}\ ,\ \pl_\m\pl_\be h_\ab\ $
, which are made from Fig.1 by connecting two suffix-lines, 
are expressed as in Fig.2.


   \begin{center}
Fig.2
\ 2-tensors of (a)\ 
$\pl^2h_\ab\ ,\ (b)\ \pl_\m\pl_\n h_{\al\al}\ $ and (c)\ $\pl_\m\pl_\be h_\ab\ $
   \end{center}


Two independent  invariants (0-tensors)\ :\ 
$P\equiv \pl_\m\pl_\m h_{\al\al},\ Q\equiv \pl_\al\pl_\be h_{\ab}\ $
, which are made from Fig.2 by connecting the remaining two suffix-lines, are
expressed as in Fig.3.


   \begin{center}
Fig.3\ Invariants of 
$P\equiv \pl_\m\pl_\m h_{\al\al}$\ and $Q\equiv \pl_\al\pl_\be h_{\ab}\ $.
   \end{center}

P and Q are all possible invariants  of $\pl\pl h$-type.
All suffix-lines of Fig.3 are closed. We easily see the following lemma
is valid.
\begin{description}
\item[Lemma 1]\ 
Generally all suffix-lines of invariants are  {\it closed}.
We call a closed suffix-line a {\it suffix-loop}.
\end{description}

\section{Representation of $(\pl\pl h)^2$-tensors and invariants}
\q Now we begin to deal with 'products' of two $\pl\pl h$-tensors.
As examples of SO(n)-tensors, we have the representations of Fig.4
for $\pl_\m\pl_\n h_\ab \pl_\m\pl_\n h_{\ga\del}$ and
$\pl_\m\pl_\n h_\ab \pl_\n\pl_\la h_{\la\be}$.

   \begin{center}
Fig.4\ 
Graphical Representations of 
$\pl_\m\pl_\n h_\ab \cdot\pl_\m\pl_\n h_{\ga\del}$ and
$\pl_\m\pl_\n h_\ab \cdot\pl_\n\pl_\la h_{\la\be}$.
   \end{center}

Before listing up all possible $(\pl\pl h)^2$-invariants,
let us state a lemma on a general SO(n)-invariant made of $s$ 
$\pl\pl h$-tensors.
\begin{description}
\item[Lemma 2]\ 
Let a general $(\pl\pl h)^s$-invariant ($s=1,2,\cdots$) has $l$ suffix-loops.
Let
each loop have $v_i$ h-vertices and $w_i$ dd-vertices ($i=1,2,\cdots, l-1,l$).
We have the following {\it necessary} conditions for $s,l,v_i\ \mbox{and } w_i$.
\begin{eqnarray}
\sum_{i=1}^{l}v_i=s\com\q \sum_{i=1}^{l}w_i=s\com\nn\\
v_i\geq 0\com\q w_i\geq 0\com\q v_i+w_i\geq 1\com\label{ddh2.1}\\
v_i\ ,\ w_i\ =0,1,2,\cdots\q ,\q l=1,2,3,\cdots,2s-1,2s\pr\nn
\end{eqnarray}
Here we may ignore the ordering of the elements\lb
 in a set
$\left\{
\left(\begin{array}{c}  v_i \\ w_i  \end{array}\right)\ ;\ 
i=1,2,\cdots,l-1,l
\right\}$
because the order can be arbitrarily changed by renumbering the suffix-loops.
\end{description}
This lemma is valid because the considered graph is made by contracting
all sufix-lines of $s$ 4-tensors of Fig.1.
We use the above Lemma  for the case $s=2$\ to list up all possible 
$(\pl\pl h)^2$-invariants.
\flushleft{(i) $l=1$}

\q For this case, we have 
\begin{eqnarray}
\left(\begin{array}{c}  v_1 \\ w_1  \end{array}\right)
=
\left(\begin{array}{c}  2 \\ 2  \end{array}\right)
\label{ddh2.2}
\end{eqnarray}
There are two ways to distribute two dd-vertices and two h-vertices on one 
suffix-loop. See Fig.5, where a small circle is used
to represent a dd-vertex explicitly.

   \begin{center}
Fig.5\ 
Two ways to distribute two dd-vertices ( small circles) and two h-vertices
(cross marks) upon one suffix-loop.
   \end{center}

\begin{description}
\item[Def 3]\ 
We call diagrams without bonds, like Fig.5, {\it bondless diagrams}. 
\end{description}
Finally,
taking account of the two bonds, we have three independent 
$(\pl\pl h)^2$-invariants for the case $l=1$. We name them $A1, A2$ and
$A3$ as shown in Fig.6.

   \begin{center}
Fig.6\ 
Three independent $(\pl\pl h)^2$-invariants for the case of one suffix-loop.
   \end{center}

\flushleft{(ii) $l=2$}

\q For this case, we have 
\begin{eqnarray}
\left\{
\left(\begin{array}{c}  v_1 \\ w_1  \end{array}\right)\ 
\left(\begin{array}{c}  v_2 \\ w_2  \end{array}\right)
\right\}
=
(a):
\left(\begin{array}{c}  2 \\ 0  \end{array}\right)\ 
\left(\begin{array}{c}  0 \\ 2  \end{array}\right)\com\ 
(b):
\left(\begin{array}{c}  1 \\ 1  \end{array}\right)\ 
\left(\begin{array}{c}  1 \\ 1  \end{array}\right)\com\nn\\ 
(c):
\left(\begin{array}{c}  1 \\ 0  \end{array}\right)\ 
\left(\begin{array}{c}  1 \\ 2  \end{array}\right)\com\ 
(d):
\left(\begin{array}{c}  0 \\ 1  \end{array}\right)\ 
\left(\begin{array}{c}  2 \\ 1  \end{array}\right)\com\label{ddh2.3} 
\end{eqnarray}
where the ordering of
$\left(\begin{array}{c}  v_1 \\ w_1  \end{array}\right)\ $ and 
$\left(\begin{array}{c}  v_2 \\ w_2  \end{array}\right)\ $
is irrelevant for the present classification as stated in Lemma 2.
\footnote{
The same treatment is adopted in the following other cases.
          }
Each one above has one bondless diagram as shown in Fig.7.

   \begin{center}
Fig.7\ 
Bondless diagrams for (\ref{ddh2.3}).
   \end{center}

Then we have 5 independent $(\pl\pl h)^2$-invariants for this case
$l=2$. We name them $B1,\,B2,\,B3,\,B4$\, and $QQ$\, as shown in Fig.8.
Among them $QQ$\, is a {\it disconnected diagram}. Fig.7b has two independent
ways to connect vertices by two bonds.

   \begin{center}
Fig.8\ 
Five independent $(\pl\pl h)^2$-invariants for the case of two suffix-loops.
   \end{center}

\flushleft{(iii) $l=3$}

\q For this case, we have 
\begin{eqnarray}
\left\{
\left(\begin{array}{c}  v_1 \\ w_1  \end{array}\right)\ 
\left(\begin{array}{c}  v_2 \\ w_2  \end{array}\right)\ 
\left(\begin{array}{c}  v_3 \\ w_3  \end{array}\right)
\right\}
=
(a):
\left(\begin{array}{c}  1 \\ 0  \end{array}\right)\ 
\left(\begin{array}{c}  1 \\ 0  \end{array}\right)\ 
\left(\begin{array}{c}  0 \\ 2  \end{array}\right)\com\nn\\
(b):
\left(\begin{array}{c}  0 \\ 1  \end{array}\right)\ 
\left(\begin{array}{c}  0 \\ 1  \end{array}\right)\ 
\left(\begin{array}{c}  2 \\ 0  \end{array}\right)\com\ 
(c):
\left(\begin{array}{c}  1 \\ 0  \end{array}\right)\ 
\left(\begin{array}{c}  0 \\ 1  \end{array}\right)\ 
\left(\begin{array}{c}  1 \\ 1  \end{array}\right)\pr\label{ddh2.4} 
\end{eqnarray}
Each one above has one bondless diagram as shown in Fig.9.

   \begin{center}
Fig.9\ 
Three bondless diagrams corresponding to (\ref{ddh2.4}).
   \end{center}

Then we have 4 independent $(\pl\pl h)^2$-invariants for the case
$l=3$. We name them $C1,\,C2,\,C3,\,$\, and $PQ$\, as shown in Fig.10.
Among them $PQ$\, is a  disconnected diagram. Fig.9c has two independent
ways to connect vertices by two bonds.

   \begin{center}
Fig.10\ 
Four independent $(\pl\pl h)^2$-invariants for the case of three suffix-loops.
   \end{center}

\flushleft{(iv) $l=4$}

\q For this case, we have 
\begin{eqnarray}
\left\{
\left(\begin{array}{c}  v_1 \\ w_1  \end{array}\right)\ 
\left(\begin{array}{c}  v_2 \\ w_2  \end{array}\right)\ 
\left(\begin{array}{c}  v_3 \\ w_3  \end{array}\right)\ 
\left(\begin{array}{c}  v_4 \\ w_4  \end{array}\right)
\right\}
=
\left(\begin{array}{c}  1 \\ 0  \end{array}\right)\ 
\left(\begin{array}{c}  1 \\ 0  \end{array}\right)\ 
\left(\begin{array}{c}  0 \\ 1  \end{array}\right)\ 
\left(\begin{array}{c}  0 \\ 1  \end{array}\right)
\pr\label{ddh2.5} 
\end{eqnarray}
This corresponds to one bondless diagram  shown in Fig.11.

   \begin{center}
Fig.11\ 
The bondless diagram corresponding to (\ref{ddh2.5}).
   \end{center}

Then we have a unique independent $(\pl\pl h)^2$-invariant (disconnected)
for the case
$l=4$. We name it $PP$\, as shown in Fig.12.

   \begin{center}
Fig.12\ 
The unique independent $(\pl\pl h)^2$-invariant 
for the case of four suffix-loops.
   \end{center}


\vs 1
\q We have obtained 
3($l=1$)+5($l=2$)+4($l=3$)+1($l=4$)=13 $(\pl\pl h)^2$-invariants from the necessary
conditions (\ref{ddh2.1}), Lemma 2.
( Among them 3 ones (QQ,PQ,PP) are disconnected.) 
Their independence is assured by their difference of the 
connectivity of suffix-lines, in other words, the topology of the graphs. 
Therefore, to conclude this section, 
we have completely listed up all independent $(\pl\pl h)^2$-invariants.
The ordinary mathematical expressions for the 13 invariants will be listed
in Table 1 of Sec.5.
In the next section, we
reprove the completeness of the above enumeration from the standpoint of
a suffix-permutation symmetry and the combinatorics among suffixes.

\section{Completeness of Graph Enumeration }
\q Let us examine the SO(n)-invariants listed in Sec.2 and Sec.3
from the viewpoint of the suffix-permutation symmetry (\ref{ddh.1}).

\flushleft{(i) $\pl\pl h$-invariants}

\q The $\pl\pl h$-invariants are obtained by contracting 4 indices
$(\m_1,\m_2,\m_3,\m_4)$ in $\pl_{\m_1}\pl_{\m_2}h_{\m_3\m_4}$.
All possible ways of contracting the four indices 
are given by the following 3 ones.
\begin{eqnarray}
a)\ \del_{\m_1\m_2}\del_{\m_3\m_4}\com\q
b)\ \del_{\m_1\m_3}\del_{\m_2\m_4}\com\q
c)\ \del_{\m_1\m_4}\del_{\m_2\m_3}
\pr\label{gc.1}
\end{eqnarray}
Due to the symmetry (\ref{ddh.1}), we see b) and c) give the same invariant Q.
\begin{description}
\item[Def 4]\ 
We generally call the number of occurrence of a covariant
(which includes the case of an invariant) C, when contracting
suffixes of a covariant C$'$ in all possible ways, a {\it weight} of C
from C$'$.
\end{description}
In the present case, P has a weight 1 and Q has a weight 2 (from 4-tensor
$\pl_\m\pl_\n h_\ab$).
We have an identity
between the number of all possible ways of suffix-contraction 
(\ref{gc.1}) and weights
of invariants.
\begin{eqnarray}
3=1(P)+2(Q)
\pr\label{gc.2}
\end{eqnarray}
A weight of an invariant shows  'degeneracy' in the contraction due to
its suffix-permutation symmetry.
The above identity shows 
the completeness of the enumeration of $\pl\pl h$-invariants from the viewpoint
of the permutation symmetry.

\flushleft{(ii) $(\pl\pl h)^2$-invariants}

\q We can do the same analysis for $(\pl\pl h)^2$-invariants. The number of all
possible contraction of 8 indices in the 8-tensor 
$\pl_{\m_1}\pl_{\m_2}h_{\m_3\m_4}\cdot\pl_{\m_5}\pl_{\m_6}h_{\m_7\m_8}$
is $7\times 5\times 3\times 1=105$. Let us take $B1$ of Fig.8c as an example
of weight calculation. See Fig. 13.
\begin{eqnarray}
\mbox{The weight of B1 from the 8-tensor}
 = 1\mbox{(weight of Fig.2b from 4-tensor $\pl\pl h$)}\nn\\
\times 4\mbox{(weight of Fig.2c from 4-tensor $\pl\pl h$)}
\times 2\mbox{(two ways of 2b-2c contraction)}       \nn\\
\times 2\mbox{(two ways of chosing 2b-bond and 2c-bond among 2 bonds)}=16
\pr\label{gc.3}
\end{eqnarray}


   \begin{center}
Fig.13\ 
Graph B1 for the weight calculation (\ref{gc.3}).
   \end{center}


Similarly we can obtain  weights for all other $(\pl\pl h)^2$-invariants
and the following identity holds true.
\begin{eqnarray}
105=7\times 5\times 3\times 1=16(A1)+16(A2)+16(A3)\nn\\
+16(B1)+16(B2)+4(B3)+4(B4)+4(QQ)\q\label{gc.4}\\
+2(C1)+2(C2)+4(C3)+4(PQ)+1(PP)\pr\nn
\end{eqnarray}
This identity clearly 
shows the completeness of the 13 $(\pl\pl h)^2$-invariants listed in Sect.3.

\q Weights, defined above, correspond to the symmetry factor or the statistical
factor in the Feynman diagram expansion of the field theory. Further the
above identity (\ref{gc.4}) reminds us of a similar one, in the graph theory,
called 'Polya's enumeration theorem'\cite{H}.

\section{Indices for Graphs} 
\q The graph representation is very useful in proving 
mathematical properties, such as completeness
and independence, of SO(n) invariants because the connectivity of
suffixes can be read in the topology of a graph. In practical calculation,
however, depicting graphs is cumbersome.
In order to specify
every graph of invariant succinctly, we present a set of indices which
shows  how  suffix-lines (suffixes) 
within one $\pl\pl h$~ or two $\pl\pl h$'s are
connected(contracted). 
In this section we characterize every independent graph of invariant 
by a set of some
indices. 
\footnote{
This approach is very contrasting with a standard one in the graph theory
where the incidence matrix or the adjacency matrix are used for
specifying a graph\cite{H,N}.
         }

\flushleft{(i)\ Number of Suffix Loops (\ul{$l$})}
\footnote{
All indices are underlined. For example, \ul{$l$}, \ul{tadpoleno},
\ul{tadtype}, \ul{bcn}, and \ul{vcn}.
          }

\q The number of suffix loops (\ul{$l$}) of a graph is a good index. In fact, 
every $\pl\pl h$-invariant is completely characterized by \ul{$l$}:\ \ 
\ul{$l$}=2 for P and \ul{$l$}=1 for Q. The index \ul{$l$} is not sufficient
to discriminate every $(\pl\pl h)^2$-invariant. We need the following ones, 
(ii) and (iii).

\flushleft{(ii)\ Number of Tadpoles (\ul{tadpoleno}) and Type of 
Tadpole (\ul{tadtype}[\ ])}

\begin{description}
\item[Def 5]\ 
We call a closed suffix-loop which has only one vertex, a {\it tadpole}.
When the vertex is dd-vertex (h-vertex),
its {\it tadpole type}, \ul{tadtype}[\ ], is defined
to be 0 (1). \ul{tadtype}[\ ] is assigned for each tadpole.
The number of tadpoles which a graph has, is called {\it tadpole number}
(\ul{tadpoleno})of the graph.
\end{description}

For example, 
in Fig.3, P has \ul{tadpoleno}=2 and \ul{tadtype}[1]=0 and 
\ul{tadtype}[2]=1
\footnote{
Final results should be independent of arbitrariness in numbering all
tadpoles in a graph.
}.
Q has \ul{tadpoleno}=0.

\flushleft{(iii)\ Bond Changing Number(\ul{bcn}[\ ]) and 
Vertex Changing Number(\ul{vcn}[\ ])}

\begin{description}
\item[Def 6]\ 
\ul{bcn}[\ ] and \ul{vcn}[\ ] are defined for each suffix-loop as follows.
When we trace a suffix-loop, starting from a vertex 
in a certain direction, we generally pass some vertices,
and  finally come back to the starting vertex. See Fig.14.
When we move, in the tracing, from a vertex to a next vertex, 
we compare the bonds to which the two vertices belong, and their vertex-types.
If the bonds are different, we set $\Del\mbox{\ul{bcn}}=1$, otherwise
$\Del\mbox{\ul{bcn}}=0$, 
If the vertex-types are different, we set $\Del\mbox{\ul{vcn}}=1$, otherwise
$\Del\mbox{\ul{vcn}}=0$. 
For $l$-th loop, we sum every number of $\Del\mbox{\ul{bcn}}$ and
$\Del\mbox{\ul{vcn}}$ while tracing the loop once, and
assign as
$\sum_{\mbox{along $l$-th loop}}\Delta \mbox{\ul{bcn}}\equiv$ \ul{bcn}[$l$],
$\sum_{\mbox{along $l$-th loop}}\Delta \mbox{\ul{vcn}}\equiv$ \ul{vcn}[$l$]
\footnote{
Final results should be independent of arbitrariness of numbering all
suffix-loops in a graph.
}.
\end{description}

   \begin{center}
Fig.14\ Explanation of \ul{bcn}[\ ] and \ul{vcn}[\ ] using Graph A2.
   \end{center}

Practically we calculate \ul{bcn}[\ ] and \ul{vcn}[\ ] as
explained in Appendix A.

\q In Table 1, we list all indices necessary
for discriminating every $(\pl\pl h)^2$-invariant completely.

\vs{10}

\newpage
\begin{tabular}{|c||c|c|c|c|c|}
\hline
Graph\ $\backslash$\ Indices
& \ul{$l$}& \ul{tadpoleno}& \ul{tadtype}[\ ]&\ul{bcn}[\ ]&\ul{vcn}[\ ]\\
\hline
        &       &            &              &              &              \\
$A1
=\pl_\si\pl_\la h_\mn\cdot\pl_\si\pl_\n h_{\m\la}$ 
        &  1    & 0          & nothing      & 4            & 2            \\
        &       &            &              &              &              \\
\hline
        &       &            &              &              &              \\
$A2
=\pl_\si\pl_\la h_{\la\m}\cdot\pl_\si\pl_\n h_{\mn}$
        & 1    & 0          & nothing      & 2            & 2            \\
        &       &            &              &              &              \\
\hline
        &       &            &              &              &              \\
$A3
=\pl_\si\pl_\la h_{\la\m}\cdot\pl_\m\pl_\n h_{\n\si}$
        & 1    & 0          & nothing      & 2            & 4            \\
        &       &            &              &              &              \\
\hline
\hline
        &       &            &              &              &              \\
$B1
=\pl_\n\pl_\la h_{\si\si}\cdot\pl_\la\pl_\m h_{\mn}$
        & 2    & 1          & 1            & $/$          & $/$          \\
        &       &            &              &              &              \\
\hline
        &       &            &              &              &              \\
$B2
=\pl^2 h_{\la\n}\cdot\pl_\la\pl_\m h_{\mn}$
        & 2    & 1          & 0            & $/$          & $/$          \\
        &       &            &              &              &              \\
\hline
        &       &            &              &  2          &  0          \\
\cline{5-6}
$B3
=\pl_\m\pl_\n h_{\la\si}\cdot\pl_\m\pl_\n h_{\ls}$
        & 2    & 0          & nothing      &   2          &  0          \\
\cline{5-6}
        &       &            &              &              &              \\
\hline
        &       &            &              &   2         &  2          \\
\cline{5-6}
$B4
=\pl_\m\pl_\n h_{\la\si}\cdot\pl_\la\pl_\si h_{\mn}$
        & 2    & 0          & nothing      &    2         &  2          \\
\cline{5-6}
        &       &            &              &              &              \\
\hline
        &       &            &              &   0         &  2           \\
\cline{5-6}
$Q^2
=(\pl_\m\pl_\n h_{\mn})^2$
        & 2     & 0          & nothing      &   0         &  2           \\
\cline{5-6}
        &       &            &              &              &               \\
\hline
\hline
        &       &            & 1             &              &                \\
$C1
=\pl_\m\pl_\n h_{\la\la}\cdot\pl_\m\pl_\n h_{\si\si}$
        & 3    & 2          & 1          &   $/$         &  $/$           \\
        &       &            &              &              &                \\
\hline
        &       &            &  0            &              &                \\
$C2
=\pl^2 h_{\mn}\cdot\pl^2 h_{\mn}$
        & 3    & 2          &   0          &   $/$         &  $/$           \\
        &       &            &              &              &                \\
\hline
        &       &            &  1            &   0         &  0           \\
\cline{4-6}
$C3
=\pl_\m\pl_\n h_{\la\la}\cdot\pl^2 h_{\mn}$
        & 3   & 2          & 0          &   0         &  0          \\
\cline{4-6}
        &       &            &              &   2         &  2          \\
\hline
        &       &            &     1         &   0         &  0           \\
\cline{4-6}
$PQ
=\pl^2 h_{\la\la}\cdot\pl_\m\pl_\n h_{\mn}$
        & 3   & 2          & 0          &   0         &  0          \\
\cline{4-6}
        &       &            &              &   0         &  2           \\
\hline
\hline
        &       &            &              &              &              \\
$P^2
=(\pl^2 h_{\la\la})^2$
        & 4  &  $/$       & $/$          & $/$          & $/$          \\
        &       &            &              &              &              \\
\hline
\multicolumn{6}{c}{\q}                                                 \\
\multicolumn{6}{c}{Table 1\ \  List of indices for all 
$(\pl\pl h)^2$-invariants. 
The symbol '$/$' means }\\
\multicolumn{6}{c}{'need not be calculated for discrimination'. 
}\\
\end{tabular}

\vs 1
The listed 13 invariants are independent each other because Table 1
clearly shows the topology of every graph is different.
\section{Application to Gravitational Theories}
Let us apply the obtained result to some simple problems. First
Riemann tensors are graphically represented as in Fig.15.


   \begin{center}
Fig.15\ Graphical representation of weak expansion of Rieman tensors .
   \end{center}

Using them, general invariants with the mass dimension (Mass)$^4$ are
expanded as in Table 2. 
\newpage

\begin{tabular}{|c||c|c|c|c|}
\hline
Graph      & $\qq\na^2R\qq$  & $\qq R^2\qq$  & $\qq R_\mn R^\mn\qq$ 
                                             &  $R_{\mn\ls}R^{\mn\ls}$  \\
\hline
$A1 $ &    $1$       & $0$            & $0$                 & $-2$       \\
$A2 $ &    $2$       &  $0$           & $\half$             & $0$        \\
$A3 $ &    $0$       &  $0$           & $\half$             & $0$        \\
$B1 $ &    $-2$      &  $0$           & $-1$                & $0$        \\
$B2 $ &    $2$       &  $0$           & $-1$                & $0$        \\
$B3 $ &$-\frac{3}{2}$&  $0$           & $0$                 & $1$        \\
$B4 $ &    $0$       &  $0$           & $0$                 & $1$        \\
$Q^2$ &    $0$       &  $1$           & $0$                 & $0$        \\
$C1 $ & $\half$      &  $0$           & $\fourth$           & $0$        \\
$C2 $ &    $-1$      &  $0$           & $\fourth$           & $0$        \\
$C3 $ &    $-1$      &  $0$           & $\half$             & $0$        \\
$PQ $ &    $0$       &  $-2$          & $0$                 & $0$        \\
$P^2$ &    $0$       &  $1$           & $0$                 & $0$        \\
\hline
\multicolumn{5}{c}{\q}                                                 \\
\multicolumn{5}{c}{Table 2\ \  Weak-Expansion of Invariants with (Mass)$^4$-Dim.
                              :\  $(\pl\pl h)^2$-Part }\\
\end{tabular}

\vs 1

The four invariants,
$\na^2R\ ,\ R^2\ ,\ R_\mn R^\mn\ $ and $R_{\mn\ls}R^{\mn\ls}$\ , 
are important in the Weyl anomaly calculation\cite{BD,II1} 
and (1-loop) counter term
calculation in 4 dim quantum gravity\cite{tHV,IO}. 
From the explicit result of
Table 2, we see the four invariants are independent as local functions
of $h_\mn(x)$, because the 13 $(\pl\pl h)^2$-invariants are
independent each other. In particular, the three 'products' of
Riemann tensors ($R^2\ ,\ R_\mn R^\mn\ ,R_{\mn\ls}R^{\mn\ls}$)
are 'orthogonal', at the leading order of weak field perturbation,
in the space 'spanned' by the 13 $(\pl\pl h)^2$-invariants. Note here
that the independence of the four invariants is proven for a general
metric $g_\mn=\del_\mn+h_\mn$.

\q The four general invariants above are independent and complete as
the Weyl anomaly terms. In the counter term calculation, however, 
we must take into
account the arbitrariness of total derivative terms, because the counter
term $\Del\Lcal$ is usually defined in the action as
\begin{eqnarray}
\intfx ~\Del\Lcal\com   \label{appli.1}
\end{eqnarray}
and fields $h_\mn(x)$ are usually assumed to damp sufficiently rapidly
at a boundary. A manifest total derivative term is $\sqg\na^2R$.
\begin{eqnarray}
\intfx \sqg\na^2R=\intfx \frac{\pl}{\pl x^\mu}(\sqg\na^\mu R)
\pr   \label{appli.2}
\end{eqnarray}
A nontrivial one is the Gauss-Bonnet topological quantity:
$R_{\mn\ab}R_{\ls\ga\del}\ep^{\mn\ls}\ep^{\ab\ga\del}/4$
$=R^2-4R_\mn R^\mn+R_{\mn\ab}R^{\mn\ab}\equiv I_{GB}^{4d}$
where $\ep^{\mn\ls}$ is the totally antisymmetric constant tensor
($\ep^{1234}=1$). From Table 2,
we obtain
\begin{eqnarray}
\sqg I_{GB}^{4d}=I_{GB}^{4d}+O'(h^3)\nn\\
=-2(A1)-2(A2)-2(A3)+4(B1)+4(B2)+(B3)+(B4)\nn\\
+(Q^2)-(C1)-(C2)-2(C3)-2(PQ)+(P^2)+O(h^3)\label{appli.3}\\
=\pl_\mu\pl_\al h_{\nu\be}\cdot\pl_\la\pl_\ga h_{\si\del}\cdot
\ep^{\mn\ls}\ep^{\ab\ga\del}+O(h^3)\nn\\
=\pl_\mu (\pl_\al h_{\nu\be}\cdot\pl_\la\pl_\ga h_{\si\del}\cdot
\ep^{\mn\ls}\ep^{\ab\ga\del})+O(h^3)\pr\nn
\end{eqnarray}
Surely $I_{GB}^{4d}$ can be expressed in a form of a total derivative term.
Therefore we can take, as the independent 1-loop counter terms in 4-dim pure
Einstein quantum gravity, 
the following two terms\cite{tHV}.
\begin{eqnarray}
R^2\com\q R_\mn R^{\mn}\pr   \label{appli.4}
\end{eqnarray}

\section{Conclusions and Discussions}
\q We have presented a graphical representation of global SO(n) tensors.
This approach allows us to systematically list up all and independent
SO(n) invariants. The completeness of the list is reassured by an identity
between a combinatoric number of suffixes and weights of listed terms
due to their suffix-permutation symmetries. Some indices, sufficient
for discriminating 
all $\pl\pl h$- and $(\pl\pl h)^2$- invariants, are given. 
They are 
useful in practical (computer) calculation. Finally we have applied
the result to some simple problems in the general relativity.

\q The present graphical representation for global SO(n) tensors
is complementary to that for
general tensors given in \cite{SI}. The latter one deals with only
general covariants, and its results are independent of  the perturbation.
In the covariant representation, however, it is difficult to prove
the independence of listed general invariants because there is no independent
'bases'. On the other hand, in the present case, although the analysis 
is based on the
weak field perturbation, we have independent 'bases'(like 13 $(\pl\pl h)^2$-
invariants) at each perturbation order. It allows us to prove independence
of listed invariants.

\q Stimulated by the duality properties of superstring theories,
anomaly structure of supergravities in higher dimensions (say, 6 dim and 10 dim)
recently becomes important. Generally
in n-dim gravity, Weyl anomaly is given by some combination of 
general invariants 
with dimension (Mass)$^{n}$ and 
$L$-loop counter-
terms are given by some combination of general invariants with dimension
(Mass)$^{n+2L-2}$. 
The present approach will be useful in those explicit calculation. 
The case for 6 dim has been analyzed in \cite{II2}.

\q Some results such as (\ref{gc.4}),(\ref{appli.3}) and Table 2
are obtained or checked by the computer calculation using a C-language program
\cite{SI96}.

\vs 2
\begin{flushleft}
{\bf Acknowledgement}
\end{flushleft}
The authors thank Prof.K.Murota (RIMS,Kyoto Univ.)
for discussions and comments about the present work. They express
gratitude to Prof. N.Nakanishi and Prof. K.Murota 
for reading the manuscript carefully. 

\vs 2
\begin{flushleft}
{\Large\bf Appendix A.\ Calculation of \ul{bcn}[\ ] and \ul{vcn}[\ ]
}
\end{flushleft}
\vs 1

\q We explain how to calculate the indices, \ul{bcn}[\ ] and \ul{vcn}[\ ] 
in the actual(computer) calculation. Let us consider a
$(\pl\pl h)^2$-invariant.
It has two bonds.
As an example, we take C1 in Fig.16.

\begin{description}
\item[Def 7]\ 
We assign i=0 for one bond and i=1 for the other. 'i' is the 
{\it bond number} and discriminates the two bonds. Next we assign j=0
for all dd-vertices and j=1 for all h-vertices. 
'j' is the {\it vertex-type number}
and discriminate the  vertex-type. Any vertex in a graph is specified
by a pair (i,j). 
\end{description}

   \begin{center}
Fig.16\ 
Bond number 'i' and vertex-type number 'j' for each vertex \nl in 
the invariant C1.
   \end{center}

\begin{description}
\item[Def 8]\ 
When we trace a suffix-line, along a loop, starting from a vertex 
(i$_0$,j$_0$) in a certain direction, we pass some vertices,
(i$_1$,j$_1$),(i$_2$,j$_2$),$\cdots$ and finally come back to the
starting vertex (i$_0$,j$_0$). 
We focus on the {\it change} of the bond number, i, and the
vertex-type number, j, when we pass from a vertex to the next vertex in 
the tracing (see Fig.17 and 14).
For $l$-th loop, we assign as
$\sum_{\mbox{along $l$-th loop}}|\Delta \mbox{i}|\equiv$ \ul{bcn}[$l$],
$\sum_{\mbox{along $l$-th loop}}|\Delta \mbox{j}|\equiv$ \ul{vcn}[$l$].
\end{description}


   \begin{center}
Fig. 17\ Change of i (bond number) and j (vertex-type number).\nl
Arrows indicate directions of tracings.
   \end{center}


\ul{bcn}[\ ] and \ul{vcn}[\ ] are listed for all $(\pl\pl h)^2$-invariants in Table 1.
\ul{bcn}[\ ] and \ul{vcn}[\ ] defined above satisfy the following
important properties.
\begin{enumerate}
\item
They do not depend on the starting vertex for tracing along a loop.
\item
They donot depend on the direction of the tracing.
\end{enumerate}

\vs 2
\begin{flushleft}
{\Large\bf Appendix B.\ 
Gauge-Fixing Condition and Graphical Rule
}
\end{flushleft}
\vs 1

\q In the text, we have not taken a gauge-fixing condition.
When we calculate a physical quantity
in the classical and quantum gravity, we sometimes need to impose the condition
on the metric $g_\mn$ for some reasons.  Firstly, in the case of quantizing
gravity itself or of solving a classical field equation with respect to
the gravity mode, 
we {\it must} impose the fixing condition in order to eliminate
the local freedom ($\ep^\m(x), \m=1,2,\cdots,n-1,n.$)
due to the general coordinate invariance (\ref{intro.2}):
$g_\mn\ra g_\mn+g_{\m\la}\na_\n\ep^\la+g_{\n\la}\na_\m\ep^\la$. 
Secondly, even when the condition is theoretically not neccesary ( such as
the  quantization on the fixed curved space, or the ordinary anomaly
calculation), the gauge-fixing is practically useful because it 
considerably reduces
the number of SO(n) invariants to be considered.

\q In the weak gravity case
$g_\mn=\del_\mn+h_\mn\ ,\ |h_\mn|\ll 1$, the condition is expressed
by $h_\mn$. Let us take a familiar gauge:
\begin{eqnarray}
\pl_\m h_\mn=\half\pl_\n h\com\q h\equiv h_{\la\la}\pr\q   \label{gf.1}
\end{eqnarray}
This condition leads to the following condition on the present
basic element $\pl_\m\pl_\n h_\ab$ .
\begin{eqnarray}
\pl_\la\pl_\m h_\mn=\half\pl_\la\pl_\n h\com\q h\equiv h_{\la\la}\pr\q
                                                               \label{gf.2}
\end{eqnarray}
This gives us a graphical rule shown in Fig.18.

   \begin{center}
Fig.18\ Graphical rule, expressing  (\ref{gf.2}), 
due to the gauge-fixing condition (\ref{gf.1}) .
   \end{center}


\q Let us see how does this rule reduce the number of independent
invariants given in the text. 
For $\pl\pl h$-invariants, we obtain the following
relation
\begin{eqnarray}
Q=\half P\pr   \label{gf.3}
\end{eqnarray}
For $(\pl\pl h)^2$-invariants, we obtain the following relations.
\begin{eqnarray}
A2=A3=\half B1=\frac{1}{4} C1\com\nn\\
B2=\half C3\com\q QQ=\half PQ=\frac{1}{4}PP\pr      \label{gf.4}
\end{eqnarray}
Therefore, in the gauge (\ref{gf.1}), we can reduce the number of independent
invariants from 2 to 1 for $\pl\pl h$-invariants (,say, $P$) and from 13 to 7
for $(\pl\pl h)^2$-invariants (,say, $A1,B3,B4,C1,C2,C3,PP$). 

\q We expect
this gauge-fixed treatment is practically very useful when a calculating
quantity is guaranteed to be gauge-invariant in advance.

\newpage
\begin{flushleft}
{\Large\bf Figure Captions}
\end{flushleft}
\begin{itemize}
\item
Fig.1\ 4-tensor $\pl_\m\pl_\n h_\ab$
\item
Fig.2
\ 2-tensors of 
$\pl^2h_\ab\ ,\ \pl_\m\pl_\n h_{\al\al}\ $ and $\pl_\m\pl_\be h_\ab\ $
\item
Fig.3\ Invariants of 
$P\equiv \pl_\m\pl_\m h_{\al\al}$\ and $Q\equiv \pl_\al\pl_\be h_{\ab}\ $.
\item
Fig.4\ 
Graphical Representations of 
$\pl_\m\pl_\n h_\ab \pl_\m\pl_\n h_{\ga\del}$ and
$\pl_\m\pl_\n h_\ab \pl_\n\pl_\la h_{\la\be}$.
\item
Fig.5\ 
Two ways to distribute two dd-vertices ( small circles) and two h-vertices
(cross marks) upon one suffix-loop.
\item
Fig.6\ 
Three independent $(\pl\pl h)^2$-invariants for the case of one suffix-loop.
\item
Fig.7\ 
Bondless diagrams for (\ref{ddh2.3}).
\item
Fig.8\ 
Five independent $(\pl\pl h)^2$-invariants for the case of two suffix-loops.
\item
Fig.9\ 
Three bondless diagrams corresponding to (\ref{ddh2.4}).
\item
Fig.10\ 
Four independent $(\pl\pl h)^2$-invariants for the case of three suffix-loops.
\item
Fig.11\ 
The bondless diagram corresponding to (\ref{ddh2.5}).
\item
Fig.12\ 
One independent $(\pl\pl h)^2$-invariant for the case of four suffix-loops.
\item
Fig.13\ 
Graph B1 for the weight calculation (\ref{gc.3}).
\item
Fig.14\ Explanation of \ul{bcn}[\ ] and \ul{vcn}[\ ] using Graph A2.
\item
Fig.15\ Graphical representation of weak expansion of Riemann tensors .
\item
Fig.16\ 
Bond number 'i' and vertex-type number 'j' for each vertex in 
the invariant C1.
\item
Fig. 17\ Change of i (bond number) and j (vertex-type number).\nl
Arrows indicate directions of tracings.
\item
Fig.18\ Graphical rule, expressing  (\ref{gf.2}), 
due to the gauge-fixing condition (\ref{gf.1}) .
\end{itemize}


\begin{thebibliography}{99}
\bibitem{EA} 
Alvarez E {1989 Quantum gravity: an introduction to some recent results
 {\it Rev.Mod.Phys.}{\bf 61} 561}
\bibitem{SI} 
Ichinose S {1995 Graphical representation of invariants and covariants in
general relativity {\it Class.Quantum Grav.}{\bf 12} 1021}
\bibitem{II1} 
Ichinose S and Ikeda N {1996 New Formulation of Anomaly, Anomaly Formula and
Graphical Representation {\it Phys.Rev.}{\bf D53} 5932}
\bibitem{II2} 
Ichinose S and Ikeda N {1996 Classification of Global SO(n)
Invariants and Independent General Invariants 
{\it Preprint of Univ. of Shizuoka} US-96-06}
\bibitem{H} 
Harary F {1969 Graph Theory (Reading-Menlo Park-Ontario: Addison-Wesley Pub.Co.)}
\bibitem{N} 
Nakanishi N {1971 Graph Theory and Feynman Integral
(New York-London-Paris: Gordon and Breach,Science Publisher)} 
\bibitem{BD} 
Birrel N D and Davies P C {1982 Quantum Fields in Curved Space (Cambridge: Cambridge Univ. Press)}
\bibitem{tHV} 
t'Hooft G and Veltman M {1974 One-loop divergences in the theory of gravitation 
{\it Ann.Inst.H.Poincar\'{e}} {\bf 20} 69}
\bibitem{IO} 
Ichinose S and Omote M {1982 Renormalization using the Background-Field Method {\it Nucl.Phys.}
{\bf B203} 221}
\bibitem{SI96} 
Ichinose S {1996 New Algorithm for Tensor Calculation in Field Theories
{\it Preprint of Univ. of Shizuoka} US-96-05}
\end{thebibliography}
\end{document}